\documentclass[10pt]{iopart}
\pdfoutput=1
\usepackage[final]{graphics}
\usepackage{iopams}
\usepackage{amssymb}
\usepackage{amsfonts}
\usepackage{epsfig}
\usepackage{graphicx}
\usepackage{color}
\usepackage{subfigure}
\usepackage{hyperref}

\begin{document}
\title[Alves and Ferreira, Eden clusters in three-dimensions and the KPZ  class]
{Eden clusters in three-dimensions and the Kardar-Parisi-Zhang universality class}
\author{Sidiney G. Alves and Silvio C. Ferreira}

\address{Departamento de F\'{\i}sica - 
Universidade Federal de Vi\c{c}osa, 36571-000, Vi\c{c}osa,
Minas Gerais, Brazil}

~\\
\address{Online at \href{http://iopscience.iop.org/1742-5468/2012/10/P10011}{stacks.iop.org/JSTAT/2012/P10011}}
\address{doi:\href{http://dx.doi.org/10.1088/1742-5468/2012/10/P10011}{10.1088/1742-5468/2012/10/P10011}}

\begin{abstract}
We present large-scale simulations of radial Eden clusters in  three-dimensions
and show that the growth exponent is in agreement with the value $\beta=0.242$ accepted for the
Kardar-Parisi-Zhang (KPZ) universality class. Our results refute a recent
assertion 
proposing that radial Eden growth in $d=3$ belongs to a universality class distinct from
KPZ. We associate the previously reported discrepancy to a slow convergence to the 
asymptotic limit. We also present the skewness and kurtosis in the roughening regime
for flat geometry in 2+1 dimensions.
\end{abstract}

\bigskip

The Kardar-Parisi-Zhang (KPZ) universality class was  introduced by 
the equation~\cite{KPZ}
\begin{equation}
 \frac{\partial h}{\partial t} = \nu \nabla^{2} h + \frac{\lambda}{2} (\nabla h)^{2} + \eta,
\label{eqKPZ}
\end{equation}
that describes the non-conservative evolution of an interface $h(x,t)$ subjected
to a white noise $\eta$. Many analytical~\cite{krugrev,SasaSpohnJsat},
numerical~\cite{Ferreira06,Alves11,Oliveira12,TakeuchiJstat} and
experimental~\cite{TakeuchiJSP12} evidences show that radial (curved) and flat
geometries have the same scaling exponents in 1+1 dimensions. 

The KPZ class is featured by an interface width $w$, defined as the standard
deviation of the distance between the interface points and the initial seed,
scaling in time as $w\sim t^{\beta}$ where $\beta$ is the growth exponent. In
$d=1+1$ dimensions the KPZ growth exponent is exactly known as
$\beta=1/3$~\cite{KPZ}. Eden model~\cite{Eden} is a standard KPZ system that
generates radial clusters with a rough surface. Off-lattice simulations of Eden
clusters immersed in $d=2$ dimensions confirmed the agreement with the KPZ 
growth exponent~\cite{Ferreira06,TakeuchiJstat}. Recently, Kuennen and
Wang~\cite{Kuennen} reported off-lattice simulations of radial Eden clusters in
$d=3$ dimensions, in which a growth exponent $\beta\approx 0.1$, much smaller
than the accepted KPZ value for flat substrates in 2+1 dimensions,
$\beta_{KPZ}\approx 0.24$~\cite{Kelling}, was observed.

\begin{figure}[t]
 \centering
\subfigure[]{\includegraphics[width=6cm]{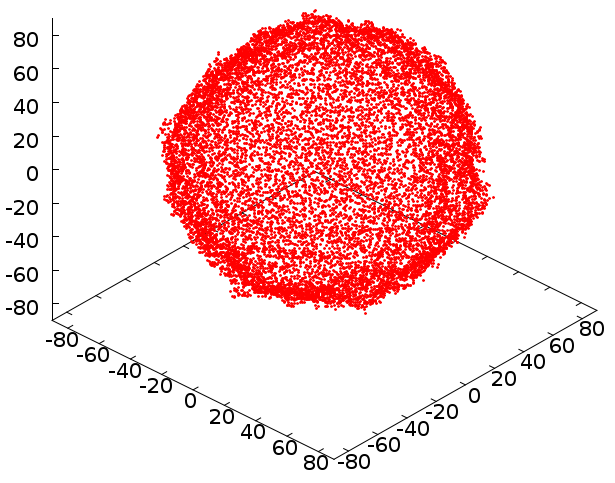}} 
\subfigure[]{\includegraphics[width=8cm]{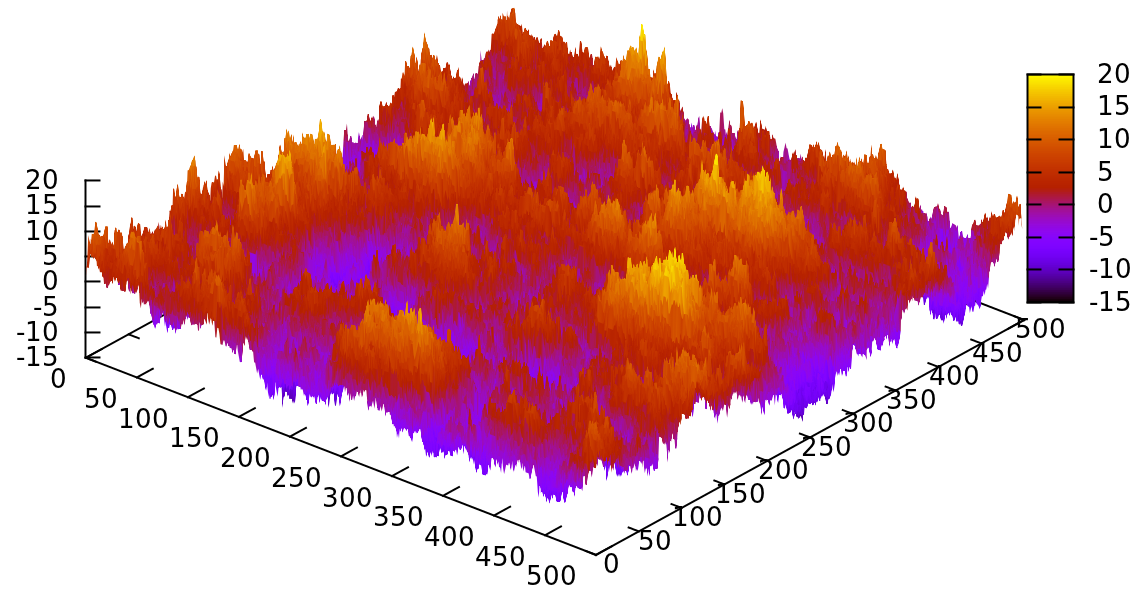}}
\caption{(a) Surface of a radial Eden~D cluster with $9\times10^5$ particles. (b)
Surface obtained with on-lattice Eden~D model after the growth of 2000 layers 
using a flat substrate of size $500\times500$.}
\label{fig:pad}
\end{figure}

In order to verify the validity of this unexpected result, we have performed
large-scale simulations of an off-lattice Eden model. The model is defined as
follows. At $t=0$, a single particle stuck to the origin is introduced. New
particles are added to the cluster, one at a time, at random positions such that
the new particle is adjacent to at least one particle in the cluster, but does
not overlap any other added particle. The dynamics runs as follows. A particle
of the cluster is randomly chosen and a random direction  is uniformly chosen on
a unitary sphere. A new particle is attached along this position if no overlap
happens. Otherwise, the simulation proceeds to the next step with the choice of
a new particle. Independently of the growth success, the time is incremented by
$\Delta t= 1/N_p$, where $N_p$ is the total number of particles in the cluster.
This version was studied in two dimensions in Ref.~\cite{TakeuchiJstat} and  we
call it of Eden~D, since there exist three other versions commonly called A,~B
and C~\cite{meakin}.  Notice that we did not care about optimizations~\cite{BJP}
to avoid uncontrolled sub-leading corrections. A typical surface of a small
cluster is shown in figure~\ref{fig:pad}(a).

\begin{figure}[ht]
 \centering
 \includegraphics[width=10cm]{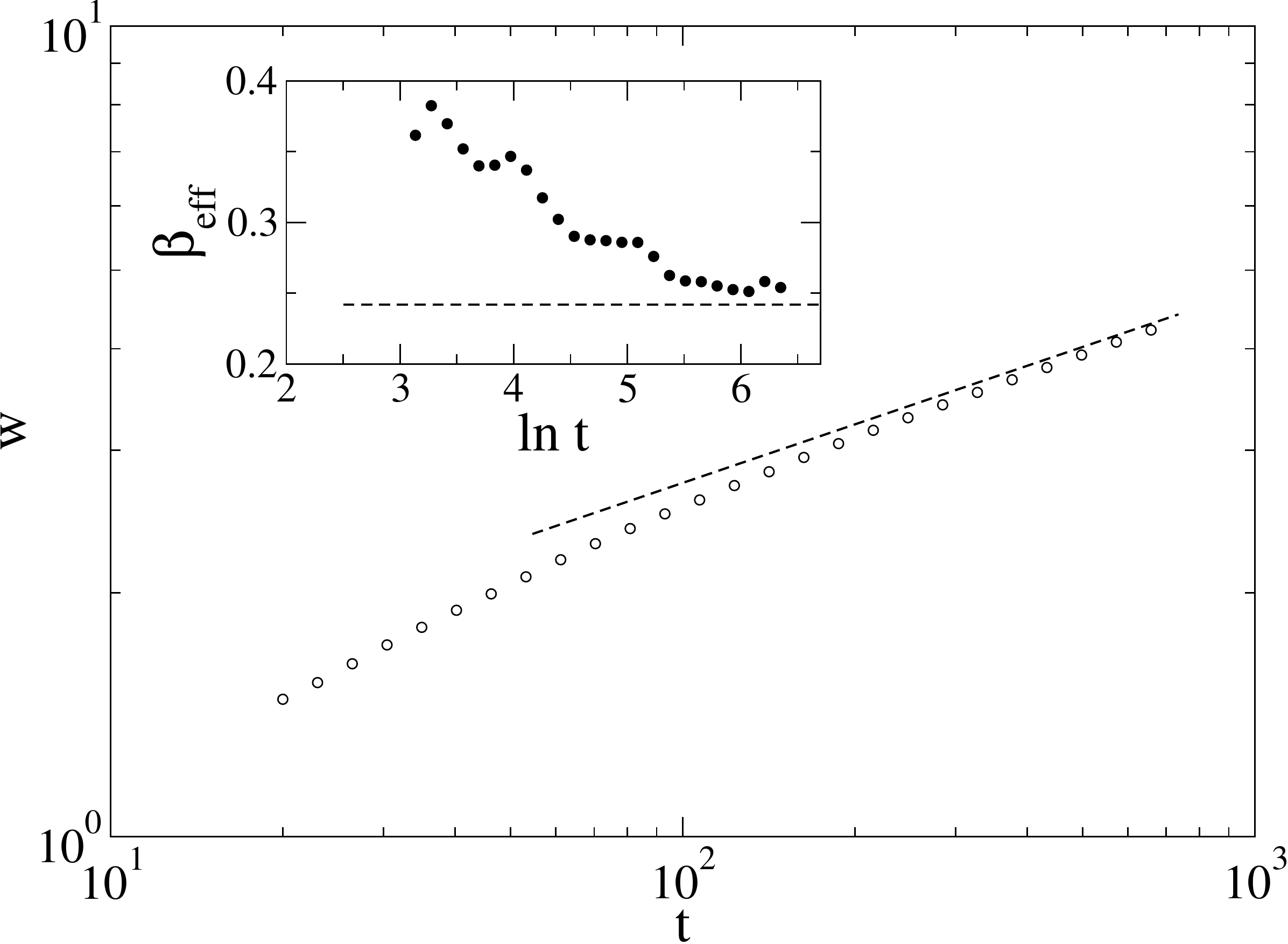}
 \caption{Main Plot: Interface width against time for off-lattice simulation of
radial Eden~D in $d=3$ dimensions. Dashed line has a slope 0.24 as a guide
to the eyes. Inset: Local slope in plot $\ln w$ vs. $\ln t$. Dashed line is the
KPZ exponent $\beta=0.24$.}
 \label{fig:off}
\end{figure}

Figure~\ref{fig:off} shows the interface width against time, averaged over 230
clusters. The largest sizes contain more than $10^8$ particles that
correspond to a size much larger than $5\times 10^5$ particles simulated in
Ref.~\cite{Kuennen}. Each sample takes about 8 hours of simulation in a Intel
Xeon 2.53 GHz CPU and 7 GB of RAM memory. The effective exponent $\beta_{eff}$
obtained from the logarithmic derivative of $w$ vs. $t$ is shown in  the inset
of figure \ref{fig:off}. The KPZ growth exponent is approached only in the limit
of very large clusters.

The model implementation in Ref.~\cite{Kuennen} is slightly different from the
Eden~D used in the present work. There, when a randomly chosen particle has
enough nearby space to add a new particle, the growth always happens. This
version is know in the literature as Eden B~\cite{Paiva07}. In Eden~D,
frustrated attempts are allowed, which, consequently, amplifies the noise and
generates larger interface fluctuations when compared to Eden~B. 

In order to shed light on this issue, we have also performed flat simulations of
both Eden models on square lattices with periodic boundary conditions along
directions $x$ and $y$. The initial condition consists in all sites occupied for
$z\le 0$. In Eden~B, particles in the interface (an occupied site with at least
one empty nearest neighbour (NN)) are randomly chosen and one of their empty
NNs, also chosen at random, is occupied. Eden~D is equal to Eden~B except that
any NN (occupied or not) can be selected. If an occupied neighbour is chosen,
the simulation runs to the next step. In both models the time step is $\Delta t
= 1/N_s$, where $N_s$ is the number of  interface sites. A typical surface for
Eden~D model is shown in figure~\ref{fig:pad}(b). 

\begin{figure}[ht]
 \centering
 \includegraphics[width=10cm]{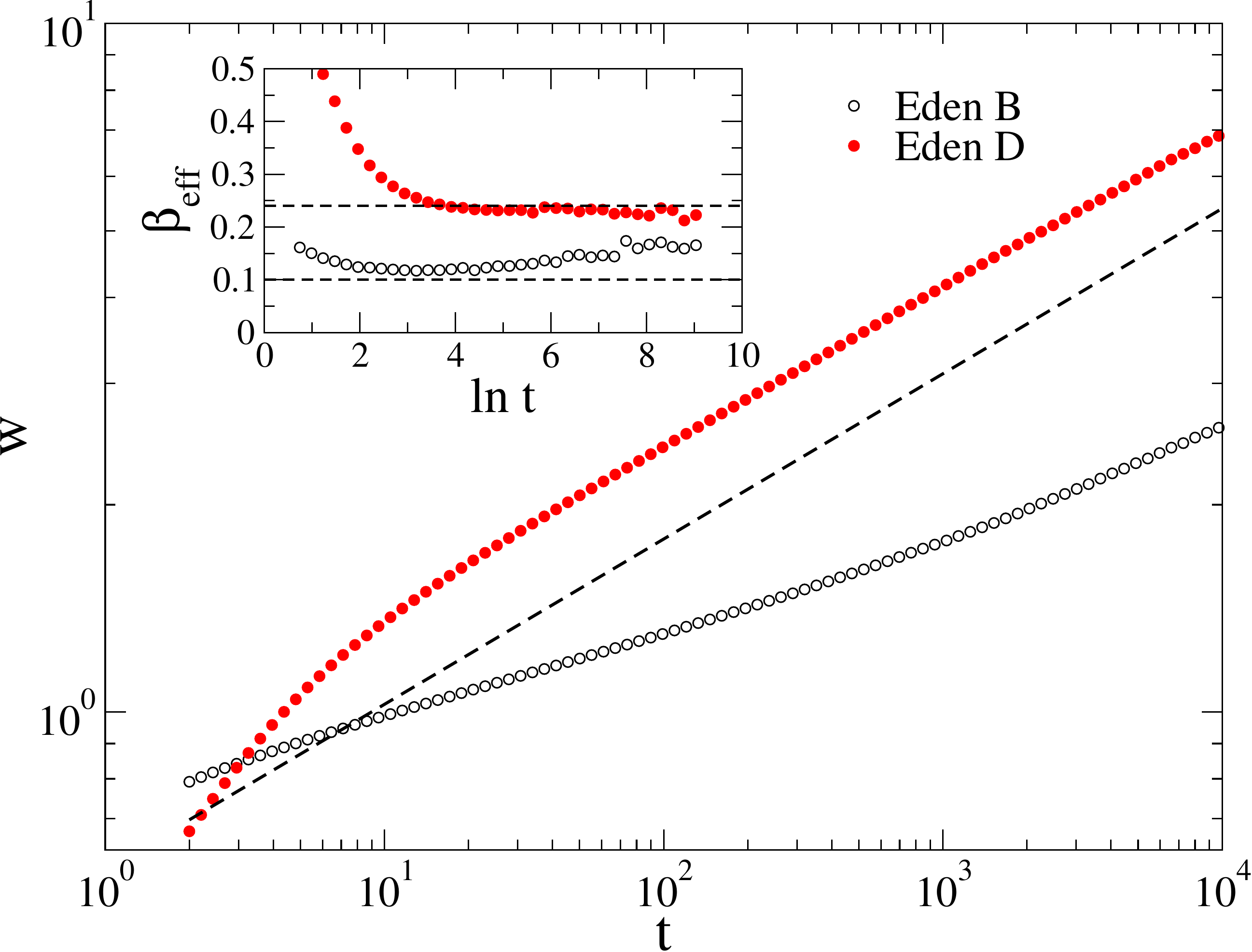}
 \caption{Main Plot: Interface width against time for lattice simulations of
Eden~B and D  on flat substrates of size
$1000\times 1000$. The dashed line has a slope 0.24 as a guide to the eyes. Inset:
Local slope in plots $\ln w$ vs. $\ln t$. Dashed lines represent the KPZ exponent
$\beta=0.24$ and the estimate of Ref.~\cite{Kuennen} $\beta=0.1$.}
 \label{fig:flat}
\end{figure}

Figure~\ref{fig:flat} shows the interface width against time for both flat models.
Averages were performed over 400 independent samples. The inset shows the
corresponding effective growth exponent. While the growth exponent in Eden~D
simulations converges quickly to a value close to the KPZ universality class, in
Eden~B, an exponent increasing from $\beta\gtrsim 0.1$ but not exhibiting a
stationary power regime is observed. These results show that Eden~B model has
very strong corrections to the scaling and that the KPZ exponent must be
observed only in exceedingly long time simulations. It is worth noticing that the
interface width in Eden~B is very small ($w<2$ lattice unities in
figure~\ref{fig:flat} and $w<1.5$ particle diameters in Ref.~\cite{Kuennen}),
which hinders the observation of the asymptotic regime. 

\begin{figure}[ht]
 \centering
 \includegraphics[width=10cm]{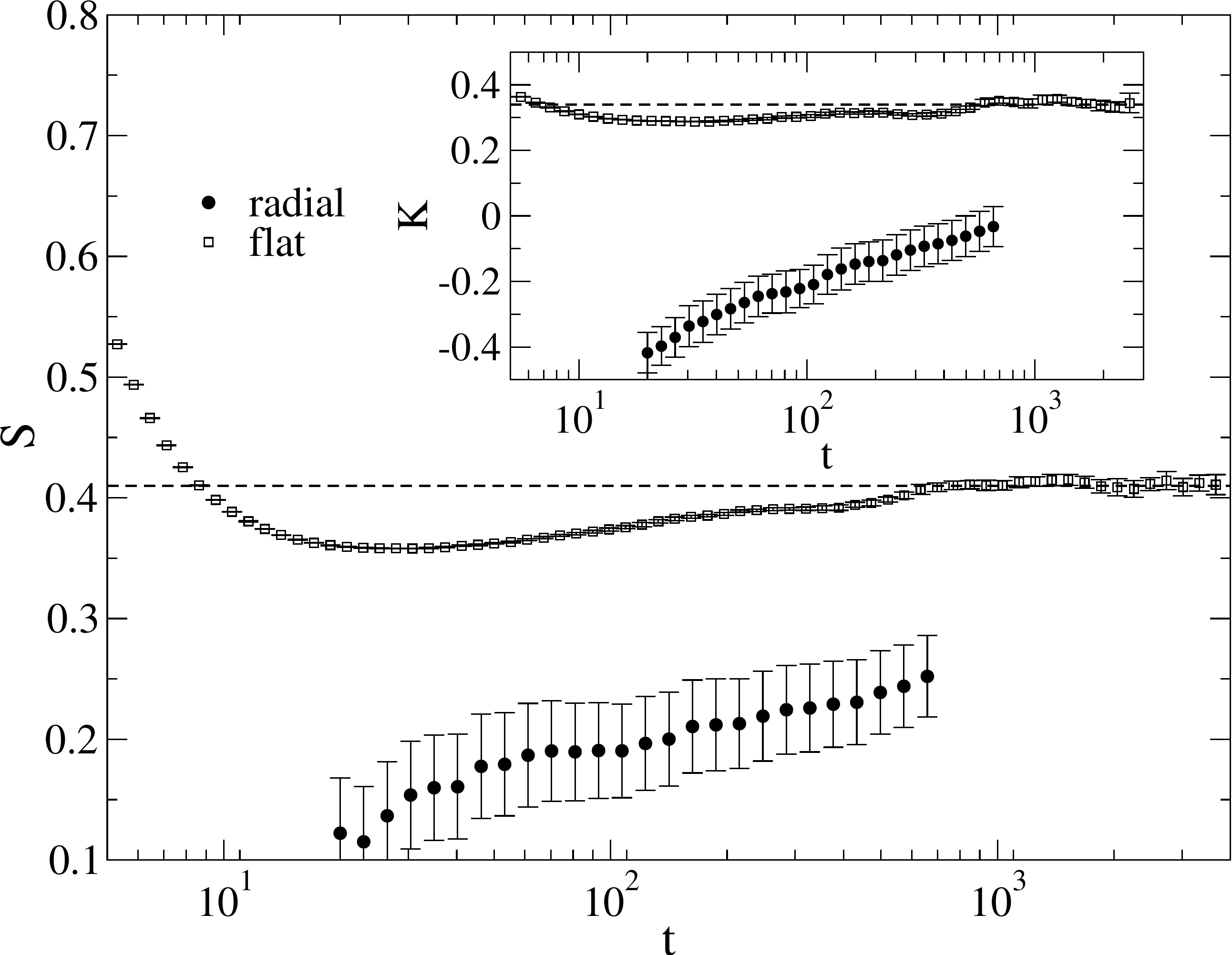}
 \caption{Main plot: Skewness against time for radial (off-lattice) and flat
(on-lattice) simulations of the Eden~D model. Flat simulations were performed in
lattices with $2200\times2200$ sites. Inset: Corresponding kurtosis against
time. Dashed lines represent the estimated cumulant ratios.}
 \label{fig:skew}
\end{figure}

Recent advances in the understanding of the KPZ universality class in 1+1
dimensions have been based on universal distributions  of the interface fluctuations
during the roughening (transient) regime~\cite{krugrev}. A basic characterization of the
distributions can be obtained via skewness $S$ and kurtosis $K$
defined as the dimensionless cumulant ratios 
\[S = \frac{\langle h ^3\rangle_c}{\langle h^2\rangle_c^{3/2}}\]
and
\[K = \frac{\langle h ^4\rangle_c}{\langle h^2\rangle_c^{2}}-3,\]
where $\langle X^n \rangle_c$ denotes the $n$th cumulant of $X$.
In our simulations, these quantities did not exhibit stationary values neither
for radial Eden~D (Fig.~\ref{fig:skew}) nor for Eden~B on flat substrates. For
Eden~D on flat substrates, the skewness and kurtosis converge to $S
= 0.411(2)$ and $K=0.34(1)$, respectively. Figure~\ref{fig:skew} shows the
skewness and kurtosis against time for radial and flat Eden~D simulations. 
It is important mentioning that cumulant ratios for Eden~D in flat simulations
have finite size effects. So, it was necessary to run shorter times and systems larger than those
used in figure~\ref{fig:flat} to obtain an accurate estimate of the cumulant ratios.
The skewness and
kurtosis obtained Eden~D agree pretty well with the ratios obtained for
other KPZ models in 2+1 dimensions~\cite{kpz3d}.
A detailed analysis of the height distributions for several KPZ models in
three-dimensional lattices will be reported elsewhere~\cite{kpz3d}.

In conclusion, we have shown that the radial Eden growth in three-dimensions
has a growth exponent in agreement with the KPZ universality class, at odds
with a previous report~\cite{Kuennen} but in consonance with the common belief.
Our results suggest that the low exponent observed in Ref.~\cite{Kuennen} comes
from the small simulated sizes and strong corrections to the scaling. 
In addition, we have presented the stationary values of the skewness and kurtosis 
during the roughening regime for flat simulations.

\bigskip

\section*{Acknowledgments}
This work was partially supported by the Brazilian
agencies CNPq and FAPEMIG.

\noindent {\small {\it Note:} After acceptance, we realized that the finite-time corrections in
the skewness and kurtosis of Eden~D model (Fig.~4) were under
estimated. Considering extrapolation to an infinite time limit, we
have reviewed our estimates and corresponding errors to  $S=0.42(1)$
and $K=0.36(2)$. Our conclusions do not change accordingly these new
estimates.}

\section*{References}

\bibliography{edenoff3d_rev}

\end{document}